\documentclass[jkps,preprint,fleqn,showpacs,showkeys]{revtex4}
\usepackage{graphicx}
\usepackage{amssymb}
\usepackage{amsmath}
\usepackage{bm}

\newcommand{\Fig}[1]{Fig.~(\ref{#1})}
\newcommand{\Eq}[1]{Eq.~(\ref{#1})}
\newcommand{\eq}[1]{Eq.~(\ref{#1})}

\newcommand{\Order}[1]{O(#1)}

\begin{document}
\setcounter{page}{0}
\title{Perturbative traveling wave solution for a flux-limited reaction-diffusion morphogenesis equation}

\author{Waipot \surname{Ngamsaad}}
\email{waipot.ng@up.ac.th}
\affiliation{Division of Physics, School of Science, University of Phayao, Phayao 56000, Thailand}

\author{Suthep \surname{Suantai}}
\affiliation{Department of Mathematics, Faculty of Science, Chiang Mai University, Chiang Mai 50200, Thailand}

\date[]{}

\begin{abstract}
In this study, we investigate a porous medium-type flux limited reaction--diffusion equation that arises in morphogenesis modeling. This nonlinear partial differential equation is an extension of the generalized Fisher--Kolmogorov--Petrovsky--Piskunov (Fisher-KPP) equation in one-dimensional space. The approximate analytical traveling wave solution is found by using a perturbation method. We show that the morphogen concentration propagates as a sharp wave front where the wave speed has a saturated value. The numerical solutions of this equation are also provided to compare them with the analytical predictions. Finally, we qualitatively compare our theoretical results with those obtained in experimental studies.
\end{abstract}


\maketitle

%%%%%%%%%%%%%%%%%%%%%%%%%%%%%%%%%%%%%%%%%%%%%%%%%%%%%%%%%%%%%%%%%%%%%%%%%%%%%%%%%%%%%%%%%%%%%%%%%%%%%%%%%%%%%%%%%%
\section{Introduction \label{sec:Intro}}
%%%%%%%%%%%%%%%%%%%%%%%%%%%%%%%%%%%%%%%%%%%%%%%%%%%%%%%%%%%%%%%%%%%%%%%%%%%%%%%%%%%%%%%%%%%%%%%%%%%%%%%%%%%%%%%%%
Reaction--diffusion models formulated using nonlinear partial differential equations have a wide range of applications in physics, chemistry, and biology (for a review, see \cite{Murray1989}). The most widely recognized reaction--diffusion model is the Fisher--Kolmogorov--Petrovsky--Piskunov (Fisher--KPP ) equation, the solution of which indicates the propagation of a traveling wave that switches between equilibrium states \cite{Fisher1937}, \cite{Tikhomirov1991}. Based on previous seminal studies, finding the traveling wave solutions of reaction--diffusion equations are attractive objectives because they can provide insights into the underlying physical dynamics of natural processes.

In developmental biology, it has been hypothesized that the concentration gradient of secreted signaling molecules known as \emph{morphogens} regulate the structure and pattern formation in tissues \cite{Dessaud2007interpretation}, \cite{Rogers2011}, \cite{Briscoe2013mechanisms}, \cite{Simon2016}. Reaction--diffusion equations have been employed as models of morphogenesis \cite{Lander2002}, \cite{Saha2006signal}, \cite{Kondo2010} since the pioneering work of Turing \cite{Turing1952chemical}, Crick \cite{Crick1970diffusion}, and Gierer and Meinhardt \cite{Gierer1972}. However, the classical theory describes the migration of morphogens as a linear diffusion process or random-walk motion from a microscopic perspective \cite{Turing1952chemical}, \cite{Crick1970diffusion}, \cite{Gierer1972}, \cite{Lander2002}, \cite{Kondo2010}. Unfortunately, experimental studies of some specific morphogens (such as Hedgehog (Hh) molecules) have shown that the classical reaction--diffusion equations are unable to capture the actual morphogenetic patterns \cite{Verbeni2013morphogenetic}, \cite{Sanchez2015modeling}. A model based on linear diffusion \cite{Saha2006signal} reproduces an unclear front, which does not agree with the experimental observations \cite{Verbeni2013morphogenetic}, \cite{Sanchez2015modeling}. In addition, the classical diffusion models has a shortcoming because it contains an infinite flux with a concentration gradient \cite{Rosenau1992Tempered}. To address various nonphysical issues, previous studies \cite{Verbeni2013morphogenetic}, \cite{Sanchez2015modeling} proposed flux-limited reaction--diffusion equations to model morphogen transport. This novel model appears to obtain more realistic morphogenetic patterns, which have been verified using experimental data \cite{Verbeni2013morphogenetic}, \cite{Sanchez2015modeling}.

The flux-limited diffusion equation can be derived using two different approaches comprising special relativistic-like mechanics \cite{Rosenau1992Tempered} and the optimal transport theory \cite{Brenier2003}. The equation was extended to the flux-limited porous medium-type diffusion equation to allow its generalization \cite{Chertock2003}, \cite{Caselles2013flux}. Together with reaction processes, the flux-limited reaction--diffusion equations have been studied widely \cite{Kurganov2006onreaction}, \cite{Andreu2010AFisher}, \cite{Andreu2012}, \cite{Garrione2015}, \cite{Campos2013ontheanalysis}, \cite{Campos2016}, \cite{Calvo2015flux}, \cite{Calvo2016pattern}, \cite{Calvo2017singular}, \cite{Calvo2018}. A one-dimensional model is a plausible representation of the real morphogenetic system, as exemplified by the propagation of Sonic Hedgehog (Shh) molecules in a neural tube along the dorsal-ventral axis \cite{Verbeni2013morphogenetic}, \cite{Sanchez2015modeling}. 

Motivated by this biological system, in the present study we investigate a one-dimensional porous medium-type flux-limited reaction--diffusion equation as a simplified model of morphogenesis. Variations in the flux-limited reaction--diffusion models have been studied previously \cite{Kurganov2006onreaction}, \cite{Andreu2010AFisher}, \cite{Campos2013ontheanalysis}, \cite{Garrione2015monotone}, \cite{Calvo2016pattern}, \cite{Campos2016}, but to the best of our knowledge, the exact solutions have not been obtained. Therefore, in the present study we aim to find the approximate analytical traveling wave solution of this equation by using a simple perturbation method, as used in the previous works \cite{Sanchezgarduno1994}, \cite{Ngamsaad2016}. This simple approximation approach is similar to asymptotic analysis \cite{Murray1989} and it uncovers two main physical features comprising the morphogen concentration profile and the propagation speed of the wave front. To obtain a precise solution, we also solve this equation using a robust fully implicit numerical scheme. Finally, we qualitatively compare our solutions with previously reported experimental evidence. We hope that our solutions provide insights into the spread and formation of patterns by morphogenesis when modeled using this simple flux-limited reaction--diffusion process. 

%In Sec. \ref{sec:Model}, we explain the derivation of the model and the approximate traveling wave solution and front speed are presented in Sec. \ref{sec:Analysis}. The numerical results are provided in In Sec. \ref{sec:Results}. And finally the conclusions are summarized in Sec. \ref{sec:Conclusions}.

%%%%%%%%%%%%%%%%%%%%%%%%%%%%%%%%%%%%%%%%%%%%%%%%%%%%%%%%%%%%%%%%%%%%%%%%%%%%%%%%%%%%%%%%%%%%%%%%%%%%%%%%%%%%%%%%%
\section{Model description \label{sec:Model}}
%%%%%%%%%%%%%%%%%%%%%%%%%%%%%%%%%%%%%%%%%%%%%%%%%%%%%%%%%%%%%%%%%%%%%%%%%%%%%%%%%%%%%%%%%%%%%%%%%%%%%%%%%%%%%%%%%
The one-dimensional porous medium-type flux-limited reaction--diffusion equation considered in this study was presented by \cite{Campos2016} and it is given by
\begin{equation}\label{eq:flux_limit_gen_FKPP}
\rho_t = \mu\left(\frac{\rho\rho_x}{\sqrt{1+\frac{\mu^2}{c_s^2}\rho_x^2}}\right)_x + R(\rho),
\end{equation}
where $\rho(x,t)$ is the morphogen concentration at position $x$ and time $t$, $\mu$ is the viscosity constant, $c_s$ is the speed of sound, and $R(\rho)$ is the reaction term \cite{Campos2016}. For the sake of simplicity, similar to previous studies \cite{Andreu2010AFisher}, \cite{Campos2013ontheanalysis}, \cite{Calvo2016pattern}, \cite{Campos2016}, the choice of the reaction term is the logistic law
\begin{equation}\label{eq:reaction}
R(\rho) = \alpha\rho\left(1-\frac{\rho}{\rho_m}\right),
\end{equation}
where $\alpha$ is the rate constant and $\rho_m$ is the maximum concentration. To capture the physical meaning of the viscosity constant, we define $\mu = c_s^2/(\gamma\rho_m)$, where $\gamma$ is the frictional rate. From \eq{eq:flux_limit_gen_FKPP} without the reaction term, the diffusion is rapid when the value of $\gamma$ is small whereas the diffusion is slow when the value of $\gamma$ is large. We rewrite \eq{eq:flux_limit_gen_FKPP} in the general form of the reaction--diffusion equation $u_t = -j_x + f(\rho)$, where $j(x,t)$ is the flux defined by $j(x,t) = \rho(x,t) V(x,t)$ and $V(x,t)$ is the velocity field. Therefore, from \eq{eq:flux_limit_gen_FKPP}, the velocity field is given by
\begin{equation}\label{eq:velocity_real}
V = -\mu\frac{\rho_x}{\sqrt{1+\frac{\mu^2}{c_s^2}\rho_x^2}}.
\end{equation}

To facilitate further analysis, we introduce the dimensionless quantities as follows: $u = \rho/\rho_m$, $t^\prime = \alpha t$, and $\epsilon = \alpha/\gamma$. Due to the constraint that $c_s$ is the highest admissible speed, we choose the dimensionless velocity as $v = V/c_s$. According to $dx^\prime \sim v dt^\prime$, the dimensionless position is provided by $x^\prime = (\alpha/c_s) x$. Now, the dimensionless concentration and velocity are limited such that $0 \leq u \leq 1$ and $0 \leq |v| \leq 1$, respectively. By substituting \eq{eq:reaction} into \eq{eq:flux_limit_gen_FKPP} with all the defined dimensionless quantities, we obtain the flux-limited reaction--diffusion equation in dimensionless form 
\begin{equation}\label{eq:flux_limit_gen_FKPP_dimensionless}
u_t = \left(\frac{\epsilon u u_x}{\sqrt{1+\epsilon^2 u_x^2}}\right)_x + u\left(1-u\right),
\end{equation}
where the prime symbols have dropped. Similarly, for \eq{eq:velocity_real}, the dimensionless velocity field is given by
\begin{equation}\label{eq:velocity_dimensionless}
v = -\frac{\epsilon u_x}{\sqrt{1+\epsilon^2 u_x^2}}.
\end{equation}
\eq{eq:flux_limit_gen_FKPP_dimensionless} is the generalized Fisher--KPP equation \cite{Newman1980} with the flux-limited diffusion extension. This equation is \emph{degenerate} at $u=0$, where it transforms from the second-order into the first-order differential equation. It is well understood that the degenerate reaction--diffusion equation produces a clear wave front interface provided that the concentration profile vanishes at a finite position \cite{Newman1980}, \cite{Murray1989}. This feature has been observed in experimental studies of morphogenesis \cite{Verbeni2013morphogenetic}, \cite{Sanchez2015modeling}. The only parameter that appears in our system is the ratio of the reaction rate relative to the frictional rate $\epsilon$, which has a crucial role in the regulation of this system. When $\epsilon \to 0$, \eq{eq:flux_limit_gen_FKPP_dimensionless} recovers a logistic reaction equation, $u_t = u\left(1-u\right)$, which has no propagating front. As $\epsilon\to\infty$, it converges to a reaction--convection equation, $u_t \approx (u u_x/|u_x|)_x + u\left(1-u\right)$, the solution of which propagates with the saturated speed $c=1$ (or $c_s$ with a physical unit). Therefore, it is likely that the flux-limited reaction--diffusion equation will eliminate the shortcoming in terms of the infinite propagation speed for the entire parameter range, or even for large concentration gradients \cite{Andreu2010AFisher}, \cite{Campos2013ontheanalysis}, \cite{Calvo2016pattern}. Due to these features, the flux-limited model is more realistic than the classical theory for describing biological transport processes.

\begin{figure}[h]
\centering\includegraphics[width=\linewidth]{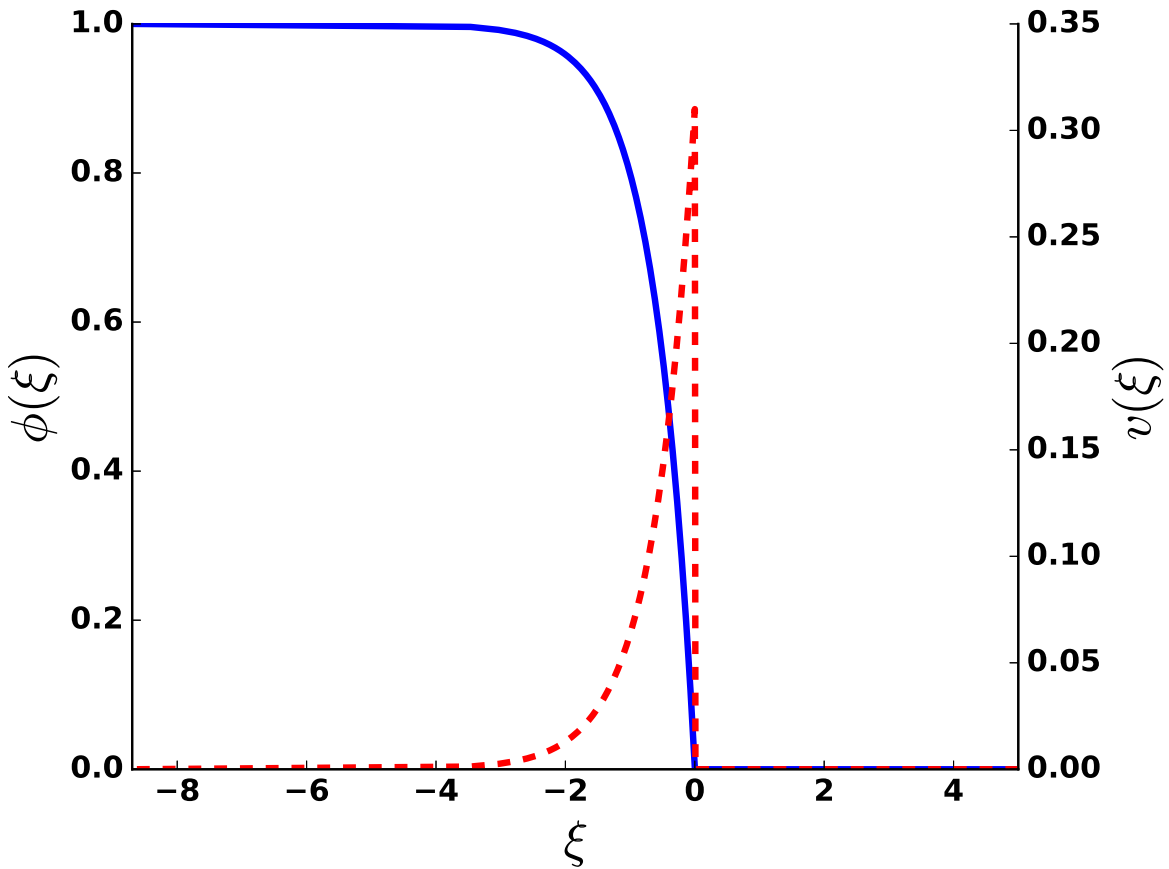}
\caption{\label{fig:analytic}
(Color online) Plots of the analytical wave profile \eq{eq:waveprofile_int} (straight line) and the corresponding velocity field \eq{eq:velocity_sol} (dashed line) for $\epsilon=0.2$.
}
\end{figure}

%%%%%%%%%%%%%%%%%%%%%%%%%%%%%%%%%%%%%%%%%%%%%%%%%%%%%%%%%%%%%%%%%%%%%%%%%%%%%%%%%%%%%%%%%%%%%%%%%%%%%%%%%%%%%%%%%
\section{Perturbative traveling wave solution \label{sec:Solution}}
%%%%%%%%%%%%%%%%%%%%%%%%%%%%%%%%%%%%%%%%%%%%%%%%%%%%%%%%%%%%%%%%%%%%%%%%%%%%%%%%%%%%%%%%%%%%%%%%%%%%%%%%%%%%%%%%%
We now assume that the solution of \eq{eq:flux_limit_gen_FKPP_dimensionless} is in the traveling wave form $u(x,t) = \phi(\xi)$, where $\xi=x-ct$ and $c$ is the speed of the front. By substituting this solution into \eq{eq:flux_limit_gen_FKPP_dimensionless} and \eq{eq:velocity_dimensionless}, respectively, we obtain
\begin{equation}\label{eq:traveling_wave1}
\left(\frac{\epsilon\phi \phi_\xi}{\sqrt{1+\epsilon^2 \phi_\xi^2}}\right)_\xi + c\phi_\xi + \phi\left(1-\phi\right) = 0,
\end{equation}
and 
\begin{equation}\label{eq:velocity_wave}
v(\xi) = -\frac{\epsilon\phi_\xi}{\sqrt{1+\epsilon^2 \phi_\xi^2}}.
\end{equation}
For simplicity, we define the rescaled variables $z = \xi/\sqrt{\epsilon}$ and $\nu = c/\sqrt{\epsilon}$ such that \eq{eq:flux_limit_gen_FKPP_dimensionless} reads
\begin{equation}\label{eq:main_PDE}
\left(\frac{\phi \phi_z}{\sqrt{1+\epsilon \phi_z^2}}\right)_z + \nu\phi_z + \phi\left(1-\phi\right) = 0.
\end{equation}
\eq{eq:main_PDE} is the main equation that we aim to analyze in this study. The exact solution of \eq{eq:main_PDE} in the general case is not available, and thus we consider a special case where $\epsilon \ll 1$, which can occur when either the growth rate is slow, $\alpha \to 0$, or the frictional rate is high, $\gamma \to \infty$. 

We employ a simple perturbation method to find the solution of \eq{eq:main_PDE}, as presented in previous studies \cite{Murray1989}, \cite{Sanchezgarduno1994}, \cite{Ngamsaad2016}. By using the Taylor expansion, \eq{eq:main_PDE} can be written in approximate form
\begin{equation}\label{eq:traveling_wave_expand}
\left[\phi\left(\phi_z-\frac{\epsilon}{2}\phi_z^3\right)\right]_z + \nu\phi_z + \phi\left(1-\phi\right) + \Order{\epsilon^2} = 0.
\end{equation}
Next, we define $\phi_z = w(\phi)$ and we then rewrite \eq{eq:traveling_wave_expand}
\begin{equation}\label{eq:traveling_wave_expand2}
\phi\left(w-\frac{3\epsilon}{2}w^3\right) w^\prime -\frac{\epsilon}{2}w^4 + w^2 + \nu w + \phi\left(1-\phi\right) = 0,
\end{equation}
where $(*)^\prime \equiv d(*)/d\phi$. The solution of \eq{eq:traveling_wave_expand2} can be written in the power series of $\epsilon$ (up to the first order)
\begin{eqnarray}
\label{eq:w_expand}
w(\phi) &=& w_0(\phi) + w_1(\phi)\epsilon + \Order{\epsilon^2}, \\
\label{eq:c_expand}
\nu &=& c_0 + c_1 \epsilon + \Order{\epsilon^2},
\end{eqnarray}
where $w_*$ and $c_*$ are the undetermined concentration wave gradients and wave speeds, respectively. After substituting \eq{eq:w_expand} and \eq{eq:c_expand} into \eq{eq:traveling_wave_expand2}, we have
\begin{eqnarray}\label{eq:traveling_wave_serie}
\lefteqn{
\phi\left(w_0 w_0^\prime  + w_0^\prime w_1\epsilon + w_0 w_1^\prime\epsilon -\frac{3}{2}w_0^3 w_0^\prime\epsilon\right)
}\nonumber\\
&& -\frac{1}{2}w_0^4\epsilon + w_0^2 + 2w_0w_1\epsilon + c_0w_0 + c_1w_0\epsilon + c_0w_1\epsilon \nonumber\\
&& + \phi\left(1-\phi\right) + \Order{\epsilon^2} = 0.
\end{eqnarray}
By comparing the coefficients of the $\epsilon^0$ and $\epsilon^1$ terms, respectively, we obtain
\begin{equation}\label{eq:zeroth_order}
\phi w_0 w_0^\prime + w_0^2 + c_0w_0 + \phi\left(1-\phi\right) = 0
\end{equation}
and 
\begin{eqnarray}\label{eq:first_order}
\lefteqn{
\phi w_0 w_1^\prime + \left(\phi w_0^\prime +2w_0 + c_0\right) w_1 
} \nonumber\\
&& -\frac{3}{2}\phi w_0^3 w_0^\prime -\frac{1}{2}w_0^4 + c_1w_0 = 0.
\end{eqnarray}
\eq{eq:zeroth_order} has known solutions in previous studies \cite{Newman1980}, \cite{Murray1989} given by 
\begin{equation}\label{eq:solution_0}
w_0(\phi) = \frac{1}{\sqrt 2}\left(\phi-1\right), \qquad c_0 = \frac{1}{\sqrt 2}.
\end{equation}
Using \eq{eq:solution_0}, \eq{eq:first_order} can be solved as shown in Appendix \ref{sec:eval}. Finally, by combining all of the terms, we have the approximate solutions (up to the first-order correction)
\begin{eqnarray}
\label{eq:w_final}
w(\phi) &=& \frac{1}{\sqrt 2}\left(\phi-1\right)\left[1+\frac{\epsilon}{6}\left(\phi^2-\frac{21}{10}\phi+\frac{6}{5}\right)\right], \\
\label{eq:c_final}
\nu &=& \frac{1}{\sqrt 2}\left(1-\frac{\epsilon}{20}\right).
\end{eqnarray}

Using the transformation $\phi_\xi = \phi_z/\sqrt{\epsilon} = w/\sqrt{\epsilon}$, from \eq{eq:velocity_wave}, we can obtain the solution for the velocity field 
\begin{equation}\label{eq:velocity_sol}
v(\phi(\xi)) = -\frac{\sqrt\epsilon w(\phi(\xi))}{\sqrt{1+\epsilon w^2(\phi(\xi))}}.
\end{equation}
In addition, from \eq{eq:w_final}, after evaluating the integral $\sqrt{\epsilon}\int d\phi/w(\phi) = \int d\xi$, we obtain the approximate analytical solution for the wave profile
\begin{eqnarray}\label{eq:waveprofile_int}
\lefteqn{
a \ln \frac{\left(\phi-1\right)^2}{1+\frac{\epsilon}{6}\left(\phi^2-\frac{21}{10}\phi+\frac{6}{5}\right)}
}\nonumber\\ 
&& + 2ab \tan^{-1} \left(b\left(20\phi-21\right)\right) + \xi_0 = \xi,
\end{eqnarray}
where $a = \frac{30\sqrt{2\epsilon}}{60+\epsilon}$, $b = \frac{\sqrt{\epsilon}}{\sqrt{2400+39\epsilon}}$ and $\xi_0 = a\left[\ln\left(1+\frac{\epsilon}{5}\right) + 2b\tan^{-1}(21b)\right]$, which is determined by using the boundary condition that $\phi(0)=0$. \eq{eq:waveprofile_int} is an implicit solution but the variables are separated explicitly, so we can plot the wave profile and the corresponding velocity field \eq{eq:velocity_sol}, as illustrated in \Fig{fig:analytic}. It should be noted that the solutions in \eq{eq:w_final}, \eq{eq:velocity_sol}, and \eq{eq:waveprofile_int} are available for $0 \leq \phi \leq 1$; otherwise, they are zero. The wave profiles have a sharp front interface where the concentration falls to zero at a finite front position. This feature is in qualitative agreement with the experimental observations because the morphogen concentration profiles have a clear invading front interface \cite{Verbeni2013morphogenetic}, \cite{Sanchez2015modeling}. 

From \eq{eq:velocity_sol}, we observe that $v(\phi) \to 1$ as $\epsilon \to \infty$ and $w(\phi) < 0$. Thus, the velocity field $v(\phi)$ in this regime is close to constant regardless of whether the solution of $w(\phi)$ is approximate. To obtain a better approximate wave speed function, we use the fact that the front speed is the velocity field at the leading edge $c=v(0)$. Thus, from \eq{eq:w_final} and \eq{eq:velocity_sol}, we have
\begin{equation}\label{eq:c_sol}
c(\epsilon) = \sqrt\frac{\epsilon }{2}\frac{1+\epsilon/5}{\sqrt{1+\frac{\epsilon}{2}\left(1+\epsilon/5\right)^2}}.
\end{equation}
By expanding \eq{eq:c_sol}, we can prove that $c/\sqrt\epsilon \approx \left(1-\frac{\epsilon}{20}\right)/\sqrt 2 + \Order{\epsilon^2} = \nu$, which is consistent with the first-order approximate solution in \eq{eq:c_final}. As $\epsilon \to \infty$, from \eq{eq:c_sol}, the wave speed reaches the limited value at $c=1$ (or $c_s$ with a physical unit), which proves that this flux-limited reaction--diffusion equation provides the saturated wave speed, as required for biological applications \cite{Verbeni2013morphogenetic}, \cite{Sanchez2015modeling}.

\begin{figure}[h]
\centering\includegraphics[width=\linewidth]{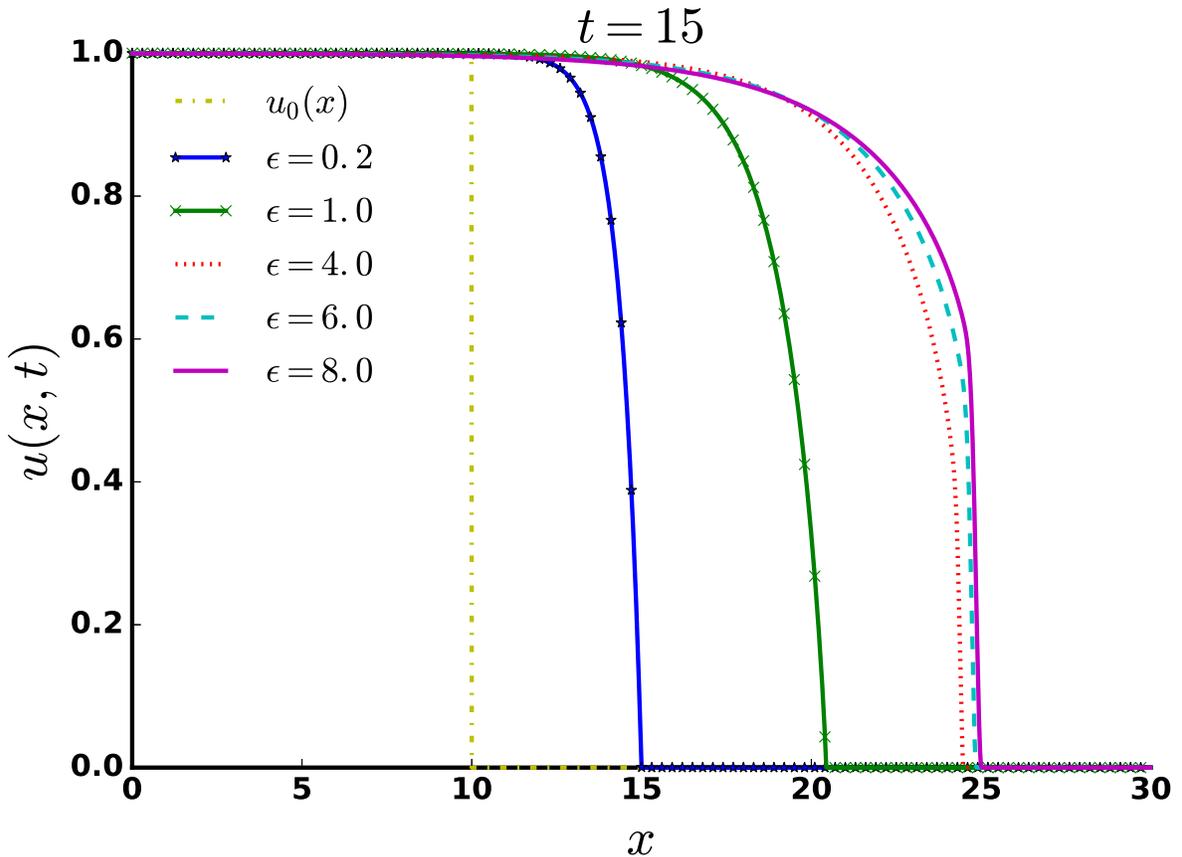}
\caption{\label{fig:num_density}
(Color online) Numerical concentration profiles at $t=15$ for selected values of the rate ratio $\epsilon$. The dash-dot line represents the initial profile. 
}
\end{figure}

\begin{figure}[h]
\centering\includegraphics[width=\linewidth]{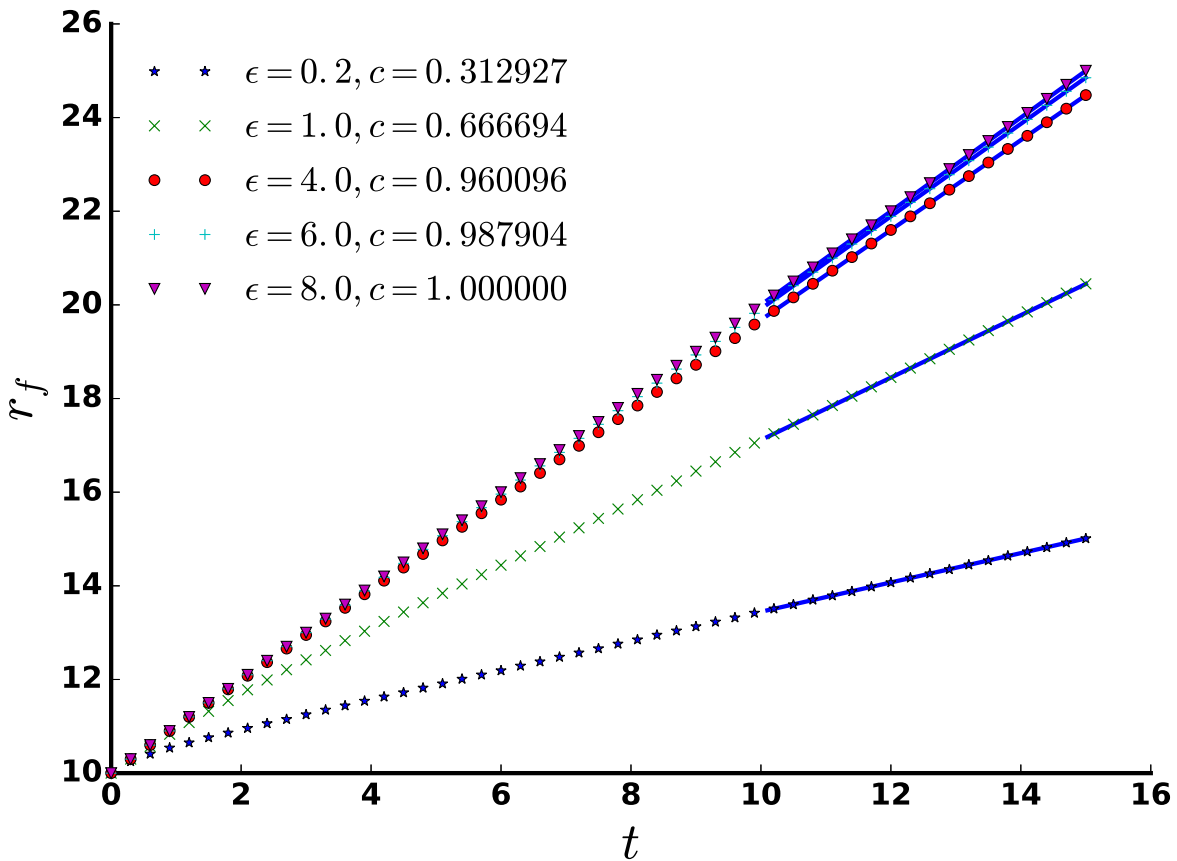}
\caption{\label{fig:num_front}
(Color online) Front position versus time corresponding to the concentration profiles in \Fig{fig:num_density}. The markers are shown for every three data points, and the solid lines represent the linear fitting curve for the last 50 data.
}
\end{figure}

%%%%%%%%%%%%%%%%%%%%%%%%%%%%%%%%%%%%%%%%%%%%%%%%%%%%%%%%%%%%%%%%%%%%%%%%%%%%%%%%%%%%%%%%%%%%%%%%%%%%%%%%%%%%%%%%%
\section{Numerical solutions and discussion \label{sec:numer}}
%%%%%%%%%%%%%%%%%%%%%%%%%%%%%%%%%%%%%%%%%%%%%%%%%%%%%%%%%%%%%%%%%%%%%%%%%%%%%%%%%%%%%%%%%%%%%%%%%%%%%%%%%%%%%%%%%
To compare the analytical predictions with more accurate numerical values, we solve the dimensionless flux-limited reaction--diffusion equation (\eq{eq:flux_limit_gen_FKPP_dimensionless}) by using a nonstandard fully implicit finite-difference method \cite{Eberl2007}, \cite{Ngamsaad2016}. First, we rewrite \eq{eq:flux_limit_gen_FKPP_dimensionless} in the usual form of the reaction--diffusion equation
\begin{equation}\label{eq:flux_limit_standard}
u_t = \left[M(u,u_x) u_x\right]_x + f(u)u,
\end{equation}
where $M(u,u_x) = \epsilon u/\sqrt{1+\epsilon^2 u_x^2}$, which is equivalent to the nonlinear diffusion coefficient, and $f(u) = 1-u$ denotes the nonlinear reaction rate. It is known that solving \eq{eq:flux_limit_standard} with a standard explicit method is inefficient due to the variable diffusion coefficient \cite{NumericalRecipes}. In addition, solving with a standard implicit scheme is even more difficult due to the nonlinearity of the equation. The idea of the nonstandard fully implicit finite-difference method is that only linear terms are discretized forward in time. Thus, we define the discrete space and time as follows: $x_i = i\delta x$, $t_n=n\delta t$, where $\delta x$ is the grid spacing, $\delta t$ is the time step, $i \in \lbrace0,1,2,\ldots,J\rbrace$, $n \in \lbrace0,1,2,\ldots,N\rbrace$, and $J$ and $N$ are integers. Now, the discrete concentration reads $u^{n}_i = u(x_i,t_n)$. Thus, \eq{eq:flux_limit_standard} in discrete form is provided by
\begin{equation}\label{eq:RD_numer1}
\frac{\partial}{\partial t} u^{n+1}_i \approx \frac{\partial}{\partial x} \left(M^n_i\frac{\partial}{\partial x}u^{n+1}_i\right) + f^n_i u^{n+1}_i,
\end{equation}
where $M^n_i = M(u^{n}_i, \partial u^{n}_i/\partial x)$ and $f^n_i = 1-u^{n}_i$. By using this approach, \eq{eq:RD_numer1} can be evaluated as a tridiagonal matrix equation in the usual manner. It has been proved that this numerical scheme is sufficiently stable for solving this type of nonlinear partial differential equation. A complete evaluation of the algorithm and its stability analysis are presented in Appendix \ref{sec:nonstandard_scheme}. 

In our computations, we set the grid spacing and the time step as $\delta x = 0.01$ and $\delta t = 0.01$, respectively. All of the calculations were performed using 3,000 grids with 1,500 iterations, which covered a spatial length of 30 and total time of 15 in dimensionless units. The initial concentration profile, $u_0(x)$, was set to a step function:
\begin{equation}\label{eq:step_fn}
u_0(x) = \left\{ 
\begin{array}{lc} 
1, & x < 10 \\ 
0, & x \geq 10.
\end{array} \right.
\end{equation}
The zero flux condition, $u_x=0$, was imposed at the boundaries. Illustrations of the concentration profiles obtained with various rate ratios ($\epsilon$) using the numerical method are shown in \Fig{fig:num_density}. We found that the concentration profiles evolved with the sharp traveling wave, which decreased to zero at a finite front position $r_f$, as predicted by the analytical solution. The wave profile had a smoother interface as the value of $\epsilon$ increased because the frictional rate $\gamma$ was small relative to the growth rate $\alpha$, so the morphogens migrated toward the free space more rapidly. For large values of $\epsilon$, the profiles obtained over an equal time tended to overlap, thereby demonstrating that the front speed tended to reach a saturated value. 

The front positions were collected every $t=0.1$ to compute the wave speed. Due to numerical deviations, the front position was determined by the first position where the concentration was lower than $1\times 10^{-6}$. The last 50 data points were selected for fitting with the linear equation, $r_f = ct + r_0$, where the wave speed was the slope of the fitted equation. Plots of the corresponding front positions versus time are shown in \Fig{fig:num_front}. Our calculated numerical front positions fitted well with the linear equation, thereby indicating that the concentration propagated with a constant front speed. 

\begin{figure}[h]
\centering\includegraphics[width=\linewidth]{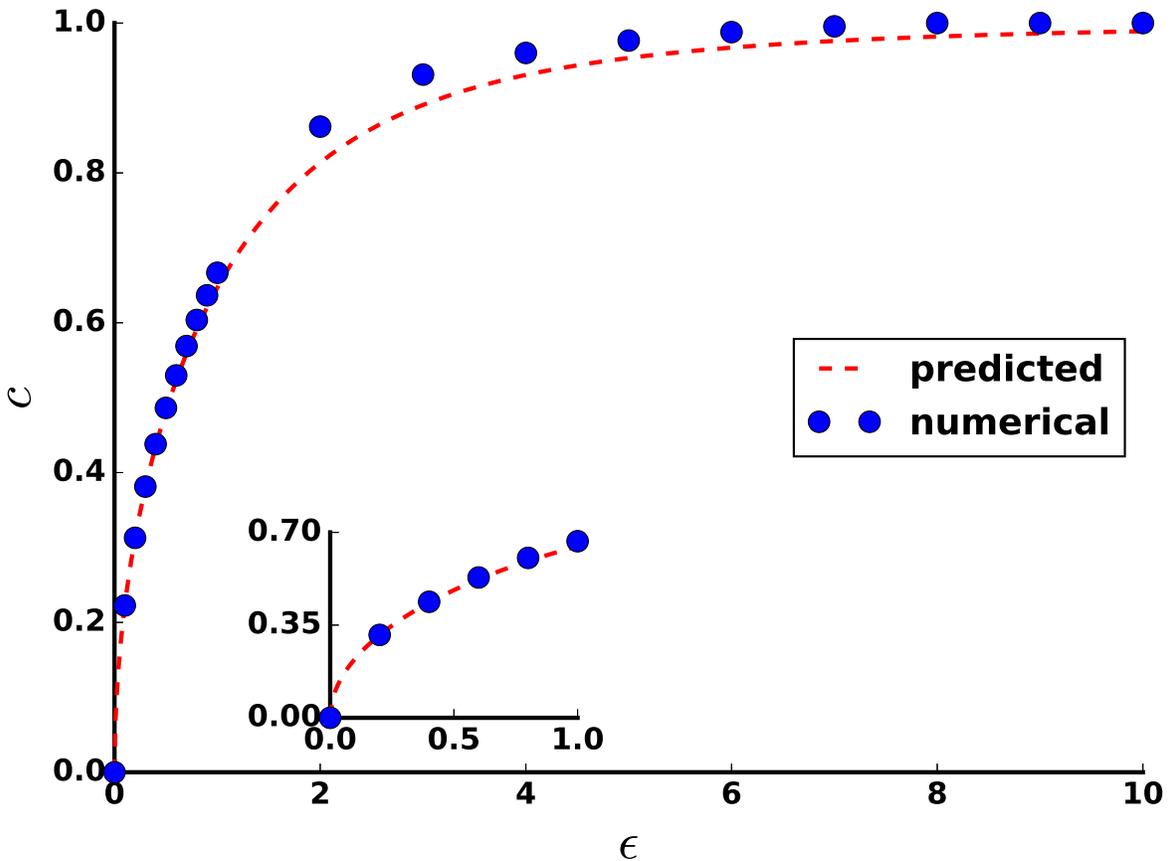}
\caption{\label{fig:front_speed}
(Color online) Front speed versus the rate ratio $\epsilon$. The circular markers represent the numerical results and the dashed lines denote the front speed predicted using \eq{eq:c_sol}. The inset shows the results for small values of $\epsilon$.
}
\end{figure}

A plot of the numerical front speed $c$ versus the rate ratio $\epsilon$ is compared with the analytical predictions obtained using \eq{eq:c_sol} in \Fig{fig:front_speed}. Both the analytical and numerical data showed that the front speed increased with $\epsilon$, and it reached a saturated value at $c=1$ as $\epsilon$ approached a large value. The analytical results agreed well with the numerical data for a small value of the rate ratio ($\epsilon \ll 1$) because the correction of our analytical solution was only $\Order{\epsilon^2}$. 

Our analytical and numerical solutions of this simple flux-limited reaction--diffusion equation capture some physical features of morphogenesis. In particular, the wave profiles have a sharp front interface where the concentration decreases to zero at a finite front position. This feature is in qualitative agreement with experimental observations because the morphogen concentration profiles have a clear invading front interface \cite{Verbeni2013morphogenetic}, \cite{Sanchez2015modeling}. Finally, we proved that this flux-limited reaction-diffusion equation provides the saturated wave speed, which is a more realistic model compared with the conventional theory \cite{Rosenau1992Tempered}. 

%%%%%%%%%%%%%%%%%%%%%%%%%%%%%%%%%%%%%%%%%%%%%%%%%%%%%%%%%%%%%%%%%%%%%%%%%%%%%%%%%%%%%%%%%%%%%%%%%%%%%%%%%%%%%%%%%
\section{Conclusions \label{sec:Conclusions}}
%%%%%%%%%%%%%%%%%%%%%%%%%%%%%%%%%%%%%%%%%%%%%%%%%%%%%%%%%%%%%%%%%%%%%%%%%%%%%%%%%%%%%%%%%%%%%%%%%%%%%%%%%%%%%%%%%
In this study, we investigated a simplified morphogenesis model governed by a porous medium-type flux-limited reaction--diffusion equation. This equation is actually an extension of the generalized Fisher--KPP equation. The approximate analytical solutions of this equation were obtained using a perturbation approach. We also solved this equation by using a nonstandard fully implicit finite-difference method in order to compare the results with the analytical predictions. The results showed that the morphogen concentration propagated as a sharp traveling wave that vanished at a finite front position and it reproduced a clear front interface. The front speed increased as the ratio of the growth rate relative to the frictional rate increased, and a saturated value was reached for a larger value of this rate ratio. We found that the flux-limited reaction--diffusion model can eliminate the shortcoming of the classical models, which yield a nonphysical infinite front speed. These features are in qualitative agreement with the experimental observations.

\begin{acknowledgments}
This research was supported by the Research Fund for DPST Graduate with First Placement (\emph{Grant no.} 28/2557) funded by The Institute for the Promotion of Teaching Science and Technology (IPST). 
\end{acknowledgments}

\appendix
%%%%%%%%%%%%%%%%%%%%%%%%%%%%%%%%%%%%%%%%%%%%%%%%%%%%%%%%%%%%%%%%%%%%%%%%%%%%%%%%%%%%%%%%%%%%%%%%%%%%%%%%%%%%%%%%%
%%%%%%%%%%%%%%%%%%%%%%%%%%%%%%%%%%%%%%%%%%%%%%%%%%%%%%%%%%%%%%%%%%%%%%%%%%%%%%%%%%%%%%%%%%%%%%%%%%%%%%%%%%%%%%%%%
\section{Evaluation of $w_1(\phi)$ and $c_1$ \label{sec:eval}}
%%%%%%%%%%%%%%%%%%%%%%%%%%%%%%%%%%%%%%%%%%%%%%%%%%%%%%%%%%%%%%%%%%%%%%%%%%%%%%%%%%%%%%%%%%%%%%%%%%%%%%%%%%%%%%%%%
By substituting \eq{eq:solution_0} into \eq{eq:first_order}, we have
\begin{eqnarray}\label{eq:first_order_subs}
\lefteqn{
\phi \left(\phi-1\right) w_1^\prime + \left(3\phi-1\right) w_1 
} \nonumber\\
&& +\left[c_1 - \frac{1}{4\sqrt 2}\left(4\phi-1\right)\left(\phi-1\right)^2\right]\left(\phi-1\right) = 0.
\end{eqnarray}
\eq{eq:first_order_subs} is a linear first-order ordinary differential equation of the form
\begin{equation}\label{eq:lin_ODE}
w_1^\prime +p(\phi)w_1 = q(\phi),
\end{equation}
where $p(\phi) = (3\phi-1)/[\phi (\phi-1)]$ and $q(\phi) = \frac{1}{4\sqrt 2}(4\phi-1)(\phi-1)^2/\phi - c_1/\phi$. The solution of \eq{eq:lin_ODE} is given by $w_1 = (C + \int I(\phi)q(\phi)d\phi)/I(\phi)$ where $I(\phi) = e^{\int p(\phi)d\phi}$ is called the integrating factor and $C$ is the integral constant \cite{Arfken1985}. After evaluation, we find that $I = \phi\left(\phi-1\right)^2$, and thus we have
\begin{eqnarray}\label{eq:w1_sol}
\lefteqn{
w_1(\phi) = \frac{1}{\phi\left(\phi-1\right)^2}\left\lbrace C + \left(\phi-1\right)^3  \right. 
}\nonumber\\ 
&& \times \left. \left[\frac{1}{6\sqrt 2}\left(\phi-1\right)^2 \left(\phi-\frac{1}{10}\right) -\frac{c_1}{3}\right]\right\rbrace.
\end{eqnarray}
To remove singularities at $\phi = 0$ and $\phi = 1$, it is necessary that $C=0$ and $c_1 = -\frac{1}{20\sqrt 2}$. Thus, the first-order concentration wave gradient is obtained by 
\begin{equation}\label{eq:w1_sol_exact}
w_1(\phi) = \frac{1}{6\sqrt 2}\left(\phi-1\right) \left(\phi^2-\frac{21}{10}\phi+\frac{6}{5}\right).
\end{equation}

%%%%%%%%%%%%%%%%%%%%%%%%%%%%%%%%%%%%%%%%%%%%%%%%%%%%%%%%%%%%%%%%%%%%%%%%%%%%%%%%%%%%%%%%%%%%%%%%%%%%%%%%%%%%%%%%%
\section{Evaluation of numerical scheme and stability analysis \label{sec:nonstandard_scheme}}
%%%%%%%%%%%%%%%%%%%%%%%%%%%%%%%%%%%%%%%%%%%%%%%%%%%%%%%%%%%%%%%%%%%%%%%%%%%%%%%%%%%%%%%%%%%%%%%%%%%%%%%%%%%%%%%%%
The differential operators in \Eq{eq:RD_numer1} were discretized further by \cite{Eberl2007}, \cite{Ngamsaad2016}, and thus we obtain
\begin{eqnarray}\label{eq:RD_numer3}
\lefteqn{
\frac{u^{n+1}_i - u^{n}_i}{\delta t} = \frac{1}{\left(\delta x\right)^2} \left[M^{n}_{i+1/2}\left(u^{n+1}_{i+1}-u^{n+1}_{i}\right) \right. 
} \nonumber \\ &&
\left. - M^{n}_{i-1/2}\left(u^{n+1}_{i}-u^{n+1}_{i-1}\right)\right] 
+ f^n_i u^{n+1}_i,
\end{eqnarray}
where 
\begin{eqnarray}
M^{n}_{i-1/2} &=& M(\frac{u^{n}_{i-1}+u^{n}_{i}}{2}, \frac{u^{n}_{i}-u^{n}_{i-1}}{\delta x}), \\
M^{n}_{i+1/2} &=& M(\frac{u^{n}_{i}+u^{n}_{i+1}}{2}, \frac{u^{n}_{i+1}-u^{n}_{i}}{\delta x}).
\end{eqnarray}
It should be noted that the correction of \eq{eq:RD_numer3} is $\Order{\delta{t}, \left(\delta{x}\right)^2}$. After rearranging \eq{eq:RD_numer3}, we have
\begin{equation}\label{eq:matrix1}
\alpha^n_i u^{n+1}_{i-1} + \theta^n_i u^{n+1}_i + \beta^n_i u^{n+1}_{i+1} = u^n_i,
\end{equation}
where 
\begin{eqnarray}
\alpha^n_i &=& -\mu M^{n}_{i-1/2}, \nonumber\\
\beta^n_i &=& -\mu M^{n}_{i+1/2}, \nonumber\\
\theta^n_i &=& 1  - \delta t f^{n}_i + \mu\left(M^{n}_{i-1/2} + M^{n}_{i+1/2}\right), \nonumber\\
\mu &=& \delta t/\left(\delta x\right)^2 .
\end{eqnarray}
By imposing the zero-flux condition at the boundary grid $\Omega$, i.e., $\left. u_x \right|_\Omega \approx \frac{u^n_{\Omega+1}-u^n_{\Omega-1}}{2\delta x} + \Order{(\delta x)^2} = 0$, we find that $u^n_{\Omega-1} = u^n_{\Omega+1}$ and $M^n_{\Omega-1/2} = M^n_{\Omega+1/2}$. According to the boundary condition, we have $\beta^n_0 = -2\mu M^{n}_{1/2}$, $\alpha^n_{J} = -2\mu M^{n}_{J-1/2}$, $\theta^n_0 = 1  - \delta t f^{n}_0 + 2\mu M^{n}_{1/2}$ and $\theta^n_J = 1  - \delta t f^{n}_J + 2\mu M^{n}_{J-1/2}$. \eq{eq:matrix1} can be written as a tridiagonal matrix equation, which can be solved numerically at each time step to obtain the numerical concentration profile $u^n_i$ \cite{NumericalRecipes}, \cite{Ngamsaad2016}. The tridiagonal matrix equation is given by
\begin{equation}\label{eq:matrix_eq}
\textbf{A}^n \cdot \textbf{U}^{n+1} = \textbf{U}^{n},
\end{equation}
where
\begin{equation}\label{eq:matrix_M}
\textbf{A}^n = \left[
\begin{array}{ccccc}
\theta^n_0 & \beta^n_0      & \cdots             &  \cdots        & 0 \\
\alpha^n_1 & \theta^n_1      & \beta^n_1          &            	   & \vdots \\
\vdots     & \ddots           & \ddots             &  \ddots        &  \vdots    \\
\vdots     &                  & \alpha^n_{J-1}   & \theta^n_{J-1}  & \beta^n_{J-1} \\
0          & \cdots	          &      \cdots       & \alpha^n_{J}  & \theta^n_{J}                                 \end{array} 
\right],
\end{equation}
and
\begin{equation}\label{eq:matrix_U}
\textbf{U}^{n} = \left[
\begin{array}{cccccc}
u^{n}_0 & u^{n}_1 & u^{n}_2 & \cdots & u^{n}_{J}
\end{array} 
\right]^{\textnormal{T}} .
\end{equation}

We analyze the stability of this numerical scheme (\eq{eq:matrix1}) by using the von Neumann approach, which assumes that
\begin{equation}\label{eq:stable1}
u^n_i = \left(\lambda\right)^n e^{\mathbf{i}ki\delta x},
\end{equation}
where $\mathbf{i} = \sqrt{-1}$, $\lambda$ represents the amplification factor and $k$ is the wave number \cite{NumericalRecipes}. By substituting \eq{eq:stable1} into \eq{eq:RD_numer3}, we have $\lambda^{-1} = 1 -\delta t f^n_i - \mu M^n_{i+1/2}\left(e^{\mathbf{i}k\delta x}-1\right) + \mu M^n_{i-1/2}\left(1-e^{-\mathbf{i}k\delta x}\right)$, which can be approximated further to obtain 
\begin{equation}\label{eq:stable2}
\lambda \approx \left[ 1 -\delta t f^n_i + 4\mu M^n_i \sin^2\left(k\delta x/2\right) + \Order{\delta x}\right]^{-1}.
\end{equation}
As $u^n_i$ increases from 0 to 1, we find that $0 \leq f^n_i \leq 1$ and $0 \leq M^n_i < \infty$. At the saturated concentration $f^n_i(u^n_i=1) = 0$, it is guaranteed that $0 < \lambda < 1$. Based on \eq{eq:stable1} and \eq{eq:stable2}, the numerical solution could converge to a finite value provided that $\delta x \ll 1$ and $\delta t \ll 1$. Therefore, the algorithm is sufficiently stable for solving this type of nonlinear partial differential equation \cite{Eberl2007}, \cite{Ngamsaad2016}.

%\section*{References}

\bibliography{DDRDE_ref}   % name your BibTeX data base

\begin{thebibliography}{36}
\expandafter\ifx\csname natexlab\endcsname\relax\def\natexlab#1{#1}\fi
\expandafter\ifx\csname bibnamefont\endcsname\relax
  \def\bibnamefont#1{#1}\fi
\expandafter\ifx\csname bibfnamefont\endcsname\relax
  \def\bibfnamefont#1{#1}\fi
\expandafter\ifx\csname citenamefont\endcsname\relax
  \def\citenamefont#1{#1}\fi
\expandafter\ifx\csname url\endcsname\relax
  \def\url#1{\texttt{#1}}\fi
\expandafter\ifx\csname urlprefix\endcsname\relax\def\urlprefix{URL }\fi
\providecommand{\bibinfo}[2]{#2}
\providecommand{\eprint}[2][]{\url{#2}}

\bibitem[{\citenamefont{Murray}(1989)}]{Murray1989}
\bibinfo{author}{\bibfnamefont{J.}~\bibnamefont{Murray}},
  \emph{\bibinfo{title}{Mathematical Biology}}
  (\bibinfo{publisher}{Springer-Verlag, New York}, \bibinfo{year}{1989}).

\bibitem[{\citenamefont{Fisher}(1937)}]{Fisher1937}
\bibinfo{author}{\bibfnamefont{R.}~\bibnamefont{Fisher}},
  \bibinfo{journal}{Ann. Eugenics} \textbf{\bibinfo{volume}{7}},
  \bibinfo{pages}{355} (\bibinfo{year}{1937}).

\bibitem[{\citenamefont{Kolmogorov et~al.}(1991)\citenamefont{Kolmogorov,
  Petrovskii, and Piscounov}}]{Tikhomirov1991}
\bibinfo{author}{\bibfnamefont{A.}~\bibnamefont{Kolmogorov}},
  \bibinfo{author}{\bibfnamefont{I.}~\bibnamefont{Petrovskii}},
  \bibnamefont{and}
  \bibinfo{author}{\bibfnamefont{N.}~\bibnamefont{Piscounov}}, in
  \emph{\bibinfo{booktitle}{Selected works of AN Kolmogorov}}, edited by
  \bibinfo{editor}{\bibfnamefont{V.}~\bibnamefont{Tikhomirov}}
  (\bibinfo{publisher}{Springer}, \bibinfo{year}{1991}), pp.
  \bibinfo{pages}{242--270}.

\bibitem[{\citenamefont{Dessaud et~al.}(2007)\citenamefont{Dessaud, Yang, Hill,
  Cox, Ulloa, Ribeiro, Mynett, Novitch, and
  Briscoe}}]{Dessaud2007interpretation}
\bibinfo{author}{\bibfnamefont{E.}~\bibnamefont{Dessaud}},
  \bibinfo{author}{\bibfnamefont{L.~L.} \bibnamefont{Yang}},
  \bibinfo{author}{\bibfnamefont{K.}~\bibnamefont{Hill}},
  \bibinfo{author}{\bibfnamefont{B.}~\bibnamefont{Cox}},
  \bibinfo{author}{\bibfnamefont{F.}~\bibnamefont{Ulloa}},
  \bibinfo{author}{\bibfnamefont{A.}~\bibnamefont{Ribeiro}},
  \bibinfo{author}{\bibfnamefont{A.}~\bibnamefont{Mynett}},
  \bibinfo{author}{\bibfnamefont{B.~G.} \bibnamefont{Novitch}},
  \bibnamefont{and} \bibinfo{author}{\bibfnamefont{J.}~\bibnamefont{Briscoe}},
  \bibinfo{journal}{Nature} \textbf{\bibinfo{volume}{450}},
  \bibinfo{pages}{717} (\bibinfo{year}{2007}).

\bibitem[{\citenamefont{Rogers and Schier}(2011)}]{Rogers2011}
\bibinfo{author}{\bibfnamefont{K.~W.} \bibnamefont{Rogers}} \bibnamefont{and}
  \bibinfo{author}{\bibfnamefont{A.~F.} \bibnamefont{Schier}},
  \bibinfo{journal}{Annu. Rev. Cell Dev. Biol.} \textbf{\bibinfo{volume}{27}},
  \bibinfo{pages}{377} (\bibinfo{year}{2011}), \bibinfo{note}{pMID: 21801015}.

\bibitem[{\citenamefont{Briscoe and Th{\'e}rond}(2013)}]{Briscoe2013mechanisms}
\bibinfo{author}{\bibfnamefont{J.}~\bibnamefont{Briscoe}} \bibnamefont{and}
  \bibinfo{author}{\bibfnamefont{P.~P.} \bibnamefont{Th{\'e}rond}},
  \bibinfo{journal}{Nat. Rev. Mol. Cell Biol.} \textbf{\bibinfo{volume}{14}},
  \bibinfo{pages}{416} (\bibinfo{year}{2013}).

\bibitem[{\citenamefont{Simon et~al.}(2016)\citenamefont{Simon,
  Aguirre-Tamaral, Aguilar, and Guerrero}}]{Simon2016}
\bibinfo{author}{\bibfnamefont{E.}~\bibnamefont{Simon}},
  \bibinfo{author}{\bibfnamefont{A.}~\bibnamefont{Aguirre-Tamaral}},
  \bibinfo{author}{\bibfnamefont{G.}~\bibnamefont{Aguilar}}, \bibnamefont{and}
  \bibinfo{author}{\bibfnamefont{I.}~\bibnamefont{Guerrero}},
  \bibinfo{journal}{J. Dev. Biol.} \textbf{\bibinfo{volume}{4}}
  (\bibinfo{year}{2016}), \urlprefix\url{http://www.mdpi.com/2221-3759/4/4/34}.

\bibitem[{\citenamefont{Lander et~al.}(2002)\citenamefont{Lander, Nie, and
  Wan}}]{Lander2002}
\bibinfo{author}{\bibfnamefont{A.~D.} \bibnamefont{Lander}},
  \bibinfo{author}{\bibfnamefont{Q.}~\bibnamefont{Nie}}, \bibnamefont{and}
  \bibinfo{author}{\bibfnamefont{F.~Y.} \bibnamefont{Wan}},
  \bibinfo{journal}{Dev. Cell} \textbf{\bibinfo{volume}{2}},
  \bibinfo{pages}{785 } (\bibinfo{year}{2002}), ISSN \bibinfo{issn}{1534-5807},
  \urlprefix\url{http://www.sciencedirect.com/science/article/pii/S153458070200179X}.

\bibitem[{\citenamefont{Saha and Schaffer}(2006)}]{Saha2006signal}
\bibinfo{author}{\bibfnamefont{K.}~\bibnamefont{Saha}} \bibnamefont{and}
  \bibinfo{author}{\bibfnamefont{D.~V.} \bibnamefont{Schaffer}},
  \bibinfo{journal}{Development} \textbf{\bibinfo{volume}{133}},
  \bibinfo{pages}{889} (\bibinfo{year}{2006}).

\bibitem[{\citenamefont{Kondo and Miura}(2010)}]{Kondo2010}
\bibinfo{author}{\bibfnamefont{S.}~\bibnamefont{Kondo}} \bibnamefont{and}
  \bibinfo{author}{\bibfnamefont{T.}~\bibnamefont{Miura}},
  \bibinfo{journal}{Science} \textbf{\bibinfo{volume}{329}},
  \bibinfo{pages}{1616} (\bibinfo{year}{2010}), ISSN \bibinfo{issn}{0036-8075},
  \urlprefix\url{https://science.sciencemag.org/content/329/5999/1616}.

\bibitem[{\citenamefont{Turing}(1952)}]{Turing1952chemical}
\bibinfo{author}{\bibfnamefont{A.~M.} \bibnamefont{Turing}},
  \bibinfo{journal}{Phil. Trans. Roy. Soc. B} \textbf{\bibinfo{volume}{237}},
  \bibinfo{pages}{37} (\bibinfo{year}{1952}).

\bibitem[{\citenamefont{Crick}(1970)}]{Crick1970diffusion}
\bibinfo{author}{\bibfnamefont{F.}~\bibnamefont{Crick}},
  \bibinfo{journal}{Nature} \textbf{\bibinfo{volume}{225}},
  \bibinfo{pages}{420} (\bibinfo{year}{1970}).

\bibitem[{\citenamefont{Gierer and Meinhardt}(1972)}]{Gierer1972}
\bibinfo{author}{\bibfnamefont{A.}~\bibnamefont{Gierer}} \bibnamefont{and}
  \bibinfo{author}{\bibfnamefont{H.}~\bibnamefont{Meinhardt}},
  \bibinfo{journal}{Kybernetik} \textbf{\bibinfo{volume}{12}},
  \bibinfo{pages}{30} (\bibinfo{year}{1972}),
  \urlprefix\url{https://doi.org/10.1007/BF00289234}.

\bibitem[{\citenamefont{Verbeni et~al.}(2013)\citenamefont{Verbeni, Sánchez,
  Mollica, Siegl-Cachedenier, Carleton, Guerrero, i~Altaba, and
  Soler}}]{Verbeni2013morphogenetic}
\bibinfo{author}{\bibfnamefont{M.}~\bibnamefont{Verbeni}},
  \bibinfo{author}{\bibfnamefont{O.}~\bibnamefont{Sánchez}},
  \bibinfo{author}{\bibfnamefont{E.}~\bibnamefont{Mollica}},
  \bibinfo{author}{\bibfnamefont{I.}~\bibnamefont{Siegl-Cachedenier}},
  \bibinfo{author}{\bibfnamefont{A.}~\bibnamefont{Carleton}},
  \bibinfo{author}{\bibfnamefont{I.}~\bibnamefont{Guerrero}},
  \bibinfo{author}{\bibfnamefont{A.~R.} \bibnamefont{i~Altaba}},
  \bibnamefont{and} \bibinfo{author}{\bibfnamefont{J.}~\bibnamefont{Soler}},
  \bibinfo{journal}{Phys. Life Rev.} \textbf{\bibinfo{volume}{10}},
  \bibinfo{pages}{457 } (\bibinfo{year}{2013}), ISSN \bibinfo{issn}{1571-0645},
  \urlprefix\url{http://www.sciencedirect.com/science/article/pii/S1571064513000833}.

\bibitem[{\citenamefont{S{\'a}nchez et~al.}(2015)\citenamefont{S{\'a}nchez,
  Calvo, Ib{\'a}{\~n}ez, Guerrero, and Soler}}]{Sanchez2015modeling}
\bibinfo{author}{\bibfnamefont{{\'O}.}~\bibnamefont{S{\'a}nchez}},
  \bibinfo{author}{\bibfnamefont{J.}~\bibnamefont{Calvo}},
  \bibinfo{author}{\bibfnamefont{C.}~\bibnamefont{Ib{\'a}{\~n}ez}},
  \bibinfo{author}{\bibfnamefont{I.}~\bibnamefont{Guerrero}}, \bibnamefont{and}
  \bibinfo{author}{\bibfnamefont{J.}~\bibnamefont{Soler}}, in
  \emph{\bibinfo{booktitle}{Hedgehog signaling protocols}}
  (\bibinfo{publisher}{Springer}, \bibinfo{year}{2015}), pp.
  \bibinfo{pages}{19--33}.

\bibitem[{\citenamefont{Rosenau}(1992)}]{Rosenau1992Tempered}
\bibinfo{author}{\bibfnamefont{P.}~\bibnamefont{Rosenau}},
  \bibinfo{journal}{Phys. Rev. A} \textbf{\bibinfo{volume}{46}},
  \bibinfo{pages}{R7371} (\bibinfo{year}{1992}),
  \urlprefix\url{https://link.aps.org/doi/10.1103/PhysRevA.46.R7371}.

\bibitem[{\citenamefont{Brenier}(2003)}]{Brenier2003}
\bibinfo{author}{\bibfnamefont{Y.}~\bibnamefont{Brenier}}, in
  \emph{\bibinfo{booktitle}{Optimal transportation and applications}}
  (\bibinfo{publisher}{Springer}, \bibinfo{year}{2003}), pp.
  \bibinfo{pages}{91--121}.

\bibitem[{\citenamefont{Chertock et~al.}(2003)\citenamefont{Chertock, Kurganov,
  and Rosenau}}]{Chertock2003}
\bibinfo{author}{\bibfnamefont{A.}~\bibnamefont{Chertock}},
  \bibinfo{author}{\bibfnamefont{A.}~\bibnamefont{Kurganov}}, \bibnamefont{and}
  \bibinfo{author}{\bibfnamefont{P.}~\bibnamefont{Rosenau}},
  \bibinfo{journal}{Nonlinearity} \textbf{\bibinfo{volume}{16}},
  \bibinfo{pages}{1875} (\bibinfo{year}{2003}),
  \urlprefix\url{http://stacks.iop.org/0951-7715/16/i=6/a=301}.

\bibitem[{\citenamefont{Caselles et~al.}(2013)}]{Caselles2013flux}
\bibinfo{author}{\bibfnamefont{V.}~\bibnamefont{Caselles}}
  \bibnamefont{et~al.}, \bibinfo{journal}{Publicacions Matem{\`a}tiques}
  \textbf{\bibinfo{volume}{57}}, \bibinfo{pages}{144} (\bibinfo{year}{2013}).

\bibitem[{\citenamefont{Kurganov and Rosenau}(2006)}]{Kurganov2006onreaction}
\bibinfo{author}{\bibfnamefont{A.}~\bibnamefont{Kurganov}} \bibnamefont{and}
  \bibinfo{author}{\bibfnamefont{P.}~\bibnamefont{Rosenau}},
  \bibinfo{journal}{Nonlinearity} \textbf{\bibinfo{volume}{19}},
  \bibinfo{pages}{171} (\bibinfo{year}{2006}),
  \urlprefix\url{http://stacks.iop.org/0951-7715/19/i=1/a=009}.

\bibitem[{\citenamefont{Andreu et~al.}(2010)\citenamefont{Andreu, Caselles, and
  Mazón}}]{Andreu2010AFisher}
\bibinfo{author}{\bibfnamefont{F.}~\bibnamefont{Andreu}},
  \bibinfo{author}{\bibfnamefont{V.}~\bibnamefont{Caselles}}, \bibnamefont{and}
  \bibinfo{author}{\bibfnamefont{J.}~\bibnamefont{Mazón}},
  \bibinfo{journal}{J. Differ. Equations} \textbf{\bibinfo{volume}{248}},
  \bibinfo{pages}{2528 } (\bibinfo{year}{2010}), ISSN
  \bibinfo{issn}{0022-0396},
  \urlprefix\url{http://www.sciencedirect.com/science/article/pii/S0022039610000124}.

\bibitem[{\citenamefont{Andreu et~al.}(2012)\citenamefont{Andreu, Calvo,
  Mazón, and Soler}}]{Andreu2012}
\bibinfo{author}{\bibfnamefont{F.}~\bibnamefont{Andreu}},
  \bibinfo{author}{\bibfnamefont{J.}~\bibnamefont{Calvo}},
  \bibinfo{author}{\bibfnamefont{J.}~\bibnamefont{Mazón}}, \bibnamefont{and}
  \bibinfo{author}{\bibfnamefont{J.}~\bibnamefont{Soler}}, \bibinfo{journal}{J.
  Differ. Equations} \textbf{\bibinfo{volume}{252}}, \bibinfo{pages}{5763 }
  (\bibinfo{year}{2012}), ISSN \bibinfo{issn}{0022-0396},
  \urlprefix\url{http://www.sciencedirect.com/science/article/pii/S0022039612000411}.

\bibitem[{\citenamefont{Garrione and
  Sanchez}(2015{\natexlab{a}})}]{Garrione2015}
\bibinfo{author}{\bibfnamefont{M.}~\bibnamefont{Garrione}} \bibnamefont{and}
  \bibinfo{author}{\bibfnamefont{L.}~\bibnamefont{Sanchez}},
  \bibinfo{journal}{Bound. Value Probl.} \textbf{\bibinfo{volume}{2015}},
  \bibinfo{pages}{45} (\bibinfo{year}{2015}{\natexlab{a}}),
  \urlprefix\url{https://doi.org/10.1186/s13661-015-0303-y}.

\bibitem[{\citenamefont{Campos et~al.}(2013)\citenamefont{Campos, Guerrero,
  Óscar Sánchez, and Soler}}]{Campos2013ontheanalysis}
\bibinfo{author}{\bibfnamefont{J.}~\bibnamefont{Campos}},
  \bibinfo{author}{\bibfnamefont{P.}~\bibnamefont{Guerrero}},
  \bibinfo{author}{\bibnamefont{Óscar Sánchez}}, \bibnamefont{and}
  \bibinfo{author}{\bibfnamefont{J.}~\bibnamefont{Soler}},
  \bibinfo{journal}{Ann. I. H. Poincar\'{e} - NA}
  \textbf{\bibinfo{volume}{30}}, \bibinfo{pages}{141 } (\bibinfo{year}{2013}),
  ISSN \bibinfo{issn}{0294-1449},
  \urlprefix\url{http://www.sciencedirect.com/science/article/pii/S0294144912000637}.

\bibitem[{\citenamefont{Campos and Soler}(2016)}]{Campos2016}
\bibinfo{author}{\bibfnamefont{J.}~\bibnamefont{Campos}} \bibnamefont{and}
  \bibinfo{author}{\bibfnamefont{J.}~\bibnamefont{Soler}},
  \bibinfo{journal}{Nonlinear Anal.} \textbf{\bibinfo{volume}{137}},
  \bibinfo{pages}{266 } (\bibinfo{year}{2016}), ISSN \bibinfo{issn}{0362-546X},
  \bibinfo{note}{nonlinear Partial Differential Equations, in honor of Juan
  Luis Vázquez for his 70th birthday},
  \urlprefix\url{http://www.sciencedirect.com/science/article/pii/S0362546X1500437X}.

\bibitem[{\citenamefont{Calvo et~al.}(2015)\citenamefont{Calvo, Campos,
  Caselles, S{\'a}nchez, and Soler}}]{Calvo2015flux}
\bibinfo{author}{\bibfnamefont{J.}~\bibnamefont{Calvo}},
  \bibinfo{author}{\bibfnamefont{J.}~\bibnamefont{Campos}},
  \bibinfo{author}{\bibfnamefont{V.}~\bibnamefont{Caselles}},
  \bibinfo{author}{\bibfnamefont{O.}~\bibnamefont{S{\'a}nchez}},
  \bibnamefont{and} \bibinfo{author}{\bibfnamefont{J.}~\bibnamefont{Soler}},
  \bibinfo{journal}{EMS Surv. Math. Sci} \textbf{\bibinfo{volume}{2}},
  \bibinfo{pages}{131} (\bibinfo{year}{2015}).

\bibitem[{\citenamefont{Calvo et~al.}(2016)\citenamefont{Calvo, Campos,
  Caselles, S{\'a}nchez, and Soler}}]{Calvo2016pattern}
\bibinfo{author}{\bibfnamefont{J.}~\bibnamefont{Calvo}},
  \bibinfo{author}{\bibfnamefont{J.}~\bibnamefont{Campos}},
  \bibinfo{author}{\bibfnamefont{V.}~\bibnamefont{Caselles}},
  \bibinfo{author}{\bibfnamefont{O.}~\bibnamefont{S{\'a}nchez}},
  \bibnamefont{and} \bibinfo{author}{\bibfnamefont{J.}~\bibnamefont{Soler}},
  \bibinfo{journal}{Invent. Math.} \textbf{\bibinfo{volume}{206}},
  \bibinfo{pages}{57} (\bibinfo{year}{2016}),
  \urlprefix\url{https://doi.org/10.1007/s00222-016-0649-5}.

\bibitem[{\citenamefont{Calvo}(2017)}]{Calvo2017singular}
\bibinfo{author}{\bibfnamefont{J.}~\bibnamefont{Calvo}}, in
  \emph{\bibinfo{booktitle}{Computational Mathematics, Numerical Analysis and
  Applications}} (\bibinfo{publisher}{Springer}, \bibinfo{year}{2017}), pp.
  \bibinfo{pages}{189--194}.

\bibitem[{\citenamefont{Calvo}(2018)}]{Calvo2018}
\bibinfo{author}{\bibfnamefont{J.}~\bibnamefont{Calvo}}, \bibinfo{journal}{SeMA
  Journal} \textbf{\bibinfo{volume}{75}}, \bibinfo{pages}{173}
  (\bibinfo{year}{2018}),
  \urlprefix\url{https://doi.org/10.1007/s40324-017-0128-y}.

\bibitem[{\citenamefont{Garrione and
  Sanchez}(2015{\natexlab{b}})}]{Garrione2015monotone}
\bibinfo{author}{\bibfnamefont{M.}~\bibnamefont{Garrione}} \bibnamefont{and}
  \bibinfo{author}{\bibfnamefont{L.}~\bibnamefont{Sanchez}},
  \bibinfo{journal}{Bound. Value Probl.} \textbf{\bibinfo{volume}{2015}},
  \bibinfo{pages}{45} (\bibinfo{year}{2015}{\natexlab{b}}), ISSN
  \bibinfo{issn}{1687-2770},
  \urlprefix\url{https://doi.org/10.1186/s13661-015-0303-y}.

\bibitem[{\citenamefont{Gardu{\~n}o and Maini}(1994)}]{Sanchezgarduno1994}
\bibinfo{author}{\bibfnamefont{F.~S.} \bibnamefont{Gardu{\~n}o}}
  \bibnamefont{and} \bibinfo{author}{\bibfnamefont{P.}~\bibnamefont{Maini}},
  \bibinfo{journal}{Appl. Math. Lett.} \textbf{\bibinfo{volume}{7}},
  \bibinfo{pages}{47 } (\bibinfo{year}{1994}), ISSN \bibinfo{issn}{0893-9659},
  \urlprefix\url{http://www.sciencedirect.com/science/article/pii/0893965994900515}.

\bibitem[{\citenamefont{Ngamsaad and Suantai}(2016)}]{Ngamsaad2016}
\bibinfo{author}{\bibfnamefont{W.}~\bibnamefont{Ngamsaad}} \bibnamefont{and}
  \bibinfo{author}{\bibfnamefont{S.}~\bibnamefont{Suantai}},
  \bibinfo{journal}{Commun. Nonlinear Sci. Numer. Simulat.}
  \textbf{\bibinfo{volume}{35}}, \bibinfo{pages}{88 } (\bibinfo{year}{2016}),
  ISSN \bibinfo{issn}{1007-5704},
  \urlprefix\url{http://www.sciencedirect.com/science/article/pii/S1007570415003718}.

\bibitem[{\citenamefont{Newman}(1980)}]{Newman1980}
\bibinfo{author}{\bibfnamefont{W.}~\bibnamefont{Newman}}, \bibinfo{journal}{J.
  Theor. Biol.} \textbf{\bibinfo{volume}{85}}, \bibinfo{pages}{325}
  (\bibinfo{year}{1980}).

\bibitem[{\citenamefont{Eberl and Demaret}(2007)}]{Eberl2007}
\bibinfo{author}{\bibfnamefont{H.~J.} \bibnamefont{Eberl}} \bibnamefont{and}
  \bibinfo{author}{\bibfnamefont{L.}~\bibnamefont{Demaret}},
  \bibinfo{journal}{Electron. J. Diff. Eqns., Conference}
  \textbf{\bibinfo{volume}{15}}, \bibinfo{pages}{77} (\bibinfo{year}{2007}).

\bibitem[{\citenamefont{Press et~al.}(1988)\citenamefont{Press, Flannery,
  Teukolsky, and Vetterling}}]{NumericalRecipes}
\bibinfo{author}{\bibfnamefont{W.~H.} \bibnamefont{Press}},
  \bibinfo{author}{\bibfnamefont{B.~P.} \bibnamefont{Flannery}},
  \bibinfo{author}{\bibfnamefont{S.~A.} \bibnamefont{Teukolsky}},
  \bibnamefont{and} \bibinfo{author}{\bibfnamefont{W.~T.}
  \bibnamefont{Vetterling}}, \emph{\bibinfo{title}{Numerical Recipes in C: The
  Art of Scientific Computing}} (\bibinfo{publisher}{Cambridge University
  Press, Cambridge}, \bibinfo{year}{1988}).

\bibitem[{\citenamefont{Arfken}(1985)}]{Arfken1985}
\bibinfo{author}{\bibfnamefont{G.~B.} \bibnamefont{Arfken}},
  \emph{\bibinfo{title}{Mathematical Methods for Physicists}}
  (\bibinfo{publisher}{Academic Press, San Diego}, \bibinfo{year}{1985}).

\end{thebibliography}

\end{document}